# Magnetic anomalies in $Gd_6Co_{1.67}Si_3$ and $Tb_6Co_{1.67}Si_3$

S. Narayana Jammalamadaka, Niharika Mohapatra, Sitikantha D Das, Kartik K Iyer and E.V. Sampathkumaran[*]

*Tata Institute of Fundamental Research, Homi Bhabha Road, Colaba, Mumbai – 400005, India*

The compounds, $Gd_6Co_{1.67}Si_3$ and $Tb_6Co_{1.67}Si_3$, recently reported to form in a $Ce_6Ni_2Si_3$-derived hexagonal structure (space group: $P6_3/m$) and to order magnetically below 295 and 190 K respectively, have been investigated by detailed magnetization (M) studies in the temperature interval 1.8-330 K as a function of magnetic field (H). The points of emphasis are: We observe multiple steps in the M(H) curve for the Tb compound at 1.8K while increasing H, but these steps do not appear in the reverse cycle of H. At higher temperatures, such steps are absent. However, this 'staircase' behavior of M(H) is not observed for the Gd compound at any temperature and the isothermal magnetization is not hysteretic unlike in Tb compound. From the M(H) data measured at close intervals of temperature, we have derived isothermal entropy change ($\Delta S$) and it is found that $\Delta S$ follows a theoretically predicted $H^{2/3}$-dependence.

1.  Introduction

Recently, the family of non-stoichiometric compounds, $R_6Co_{1.67}Si_3$ (R= Rare-earth), has been reported to form [1, 2] with a hexagonal structure derived from $Ce_6Ni_2Si_3$ (space group: $P6_3/m$). The crystallographic details have been presented in great detail in the recent literature [1-5]. In this crystal structure, there are two sites for R, while Co has 3 sites. This structure consists of chains made up of trigonal prisms of R atoms sharing a face. It appears that there is a strain in the lattice, resulting in a small $c$-axis, and the partial occupation of one of the Co sites is presumably a manifestation of this strain. In this article, we focus our attention on Gd and Tb analogues, which have been found to order magnetically below 295 and 190 K respectively [1, 4]. This investigation is a continuation of our recent work [6] on hexagonal $Gd_4Co_3$, the crystal structure of which is closely related [4] to that of the family under discussion. Following initial reports on the complex magnetic behavior of these two compounds [1, 4], we considered it important to probe magnetization (M) behavior in detail as a function of temperature (T) and magnetic-field (H). We find that the features in the vicinity of the onset of magnetic ordering are sensitive to initial applications of field. The most noteworthy finding is that, at 1.8 K, there are multiple steps in the isothermal M data of Tb compound, whereas for a marginal increase of temperature, such a staircase-like behavior is absent; also, the Gd compound does not show such an anomaly. Since we measured isothermal M at several temperatures, we have also looked at the magnetocaloric effect (MCE) considering current interests to look for materials exhibiting large MCE in different temperature ranges and to look for a relationship between MCE and H. We would like to mention that, at the time of writing this article, we came to know of another article [5] on MCE behavior of the Gd compound, the results of which are in good agreement with that of ours.

2.  Experimental details

The samples in the polycrystalline form have been prepared by arc melting together stoichiometric amounts of high purity (>99.9 wt %) constituent elements in an atmosphere of argon. The weight loss after several melting was negligible. The ingots were annealed at 1073 K for 1 month in an evacuated sealed quartz tube. The samples were characterized by x-ray diffraction (Cu Kα) and patterns revealed the formation of the compounds; for the Tb compound, however, there are a few weak extra lines (about 5% intensity) as shown by asterisk in figure 1 attributable to $Tb_5Si_3$, as known in the literature [1]. The dc M(T) (T= 1.8-300 K) behavior was measured with a commercial (Quantum Design) SQUID magnetometer in the presence of different fields for the zero-field-cooled (ZFC, from 330 K for Gd sample, and from 250 K for Tb sample) and field-cooled (FC) conditions of the specimens in the form of ingots; in addition, isothermal M behavior at several temperatures was performed up to 70 kOe for the ZFC condition with the same instrument. For the Tb case, we have extended M(H) measurements to higher fields (120 kOe) with a vibrating sample magnetometer (VSM) (Oxford Instruments). In addition, for the Tb compound, we have measured ac susceptibility ($\chi$) with four different frequencies and ac field of 1 Oe to look for possible spin-glass features.

3.  Results and discussion

M(T) curves obtained in various fields are shown in figures 2 and 3 for these compounds. There is a distinct evidence for the onset of magnetic ordering near 300 K and 200 K for Gd and Tb compounds respectively. It is to be noted that, for the Gd sample, the jump in the magnetization is quite sharp and there is a weak peak at the magnetic transition, if the magnetization is measured in very low fields (see the curve for H= 50 in the inset of figure 2). Well below the magnetic transition temperature, the value of M is nearly constant. We do not find any bifurcation of ZFC and FC curves for this compound. The feature at the transition point is sharp for Tb sample as well at low fields (see figure 3). However, the ZFC-curves bifurcate presumably due to domain wall pinning effects, often reported (see, for instance, Ref. 7) in the literature for materials with long range magnetic order (but such an effect is absent for the Gd case as inferred from the data discussed above). The bifurcation temperatures (for Tb sample) are about 180 K and 30 K for H= 100 Oe and 5 kOe respectively; as noted in Ref. 1, it appears that there is another feature around 20-40 K in the form of a drop in the ZFC-curves. It is not clear whether this temperature range marks the onset of a highly metastable state, the magnetization of which is sensitive to the history of measurements. In order to address whether the ZFC-FC $\chi$ irreversibility for the Tb case arises from spin-glass behavior, we have performed ac $\chi$ measurements with 1.3, 13, 133 and 1333 Hz and we did not find any frequency dependence of the features thereby ruling out spin-glass freezing in the temperature range of investigation; there are peaks near 120, 170 and 183 K, as though there is more than one magnetic transition (see figure 3 for a typical curve). There appears to be corresponding features in the ZFC-curves, though broad, at 120 and 170 K for low dc fields in the dc $\chi$ data. One can not rule out the possibility that the weak 120K-transition arises from the $Tb_5Si_3$ secondary phase (and possibly from traces of TbCoSi) ordering magnetically in that range [8, 9]. (On the basis of M(H) data discussed below, we infer that there is an antiferromagnetic component in zero-field, though the observation of hysteretic M(H) behavior, however weak it may be at higher temperatures, implies that the net magnetic structure is ferrimagnetic, as inferred in Ref. 1). With respect to paramagnetic data, we could not look for the Curie-Weiss region for the Gd sample, as the magnetic transition is near the extreme high temperature limit of our magnetization measurements. For Tb sample, magnetic susceptibility follows Curie-Weiss behavior above 205 K as shown in the inset of figure 3 and the paramagnetic Curie temperature is about 195 K; the effective moment obtained from the linear region is ~ 9.62 $\mu_B$/Tb, which is close to the value expected for free trivalent Tb ions within the limits of experimental error (0.1 $\mu_B$).

In figures 4, we show the isothermal magnetization behavior measured at close intervals of temperature for the Gd compound. For the Gd case, there is a deviation from linearity at high fields in the paramagnetic state, attributable to short-range correlations. In the magnetically ordered state, after an initial rise (till about 5 kOe), M tends towards saturation at higher fields with a very weak gradual increase beyond 5 kOe. The curves for this compound are featureless at high fields (and hence not shown in figure 4 beyond 20 kOe). The value of the saturation moment obtained by extrapolation of the high-field data to zero field is about 6.3 $\mu_B$/Gd at 5 K, which is less than the free ion value of 7 $\mu_B$/Gd. While, if Ref. 4, a higher value of about 6.8 $\mu_B$/Gd has been reported, our value is the same as that in another recent report [10] which appeared in print after submission of this article for publication. Therefore, we conclude that there is an induced moment on Co, coupled antiparallel to that of Gd moment and/or canted antiferromagnetism. The M(H) curves are not found to be hysteretic for the Gd compound. These findings, in broad agreement with Refs. 4 and 5, imply that this compound can be classified as a soft ferrimagnet.

For the Tb sample, the curves (see figure 5) are essentially linear in the paramagnetic state (as expected). The behavior in the magnetically ordered state is complex. There is an irreversibility of M(H) curves, the degree of which gradually increases with decreasing temperature as illustrated in figure 6 for 80 and 5 K. The fact that there is a hysteresis loop, however weak it may be, even at 80 K implies that there is a ferromagnetic component even at this temperature. The value of the coercive field is small (~200 Oe), say at 80 K, and it increases with decreasing temperature, say, to about 8 kOe at 5 K. Above 20 K (in the magnetically ordered state), there is a sharp rise of M at low fields (see the curve for 80 K in figure 6) followed by a tendency towards saturation at higher fields as though there is a strong ferromagnetic component. Below 20 K, in the zero-field (virgin) state, there is a knee near zero field (see 5K-curve in figure 6). Possible presence of an antiferromagnetic component at all temperatures is supported by a weak gradual increase of M even at high fields (see figure 5) without any evidence for saturation. The value of the extrapolated saturation moment is (5 $\mu_B$/Tb) at 1.8 K, obtained by linear extrapolation of the high field data to zero field, is much less than that of fully degenerate trivalent Tb ion (9 $\mu_B$). Though, for Tb, crystal-field effect also contributes to a reduction of the saturation moment, considering similarities with the Gd compound, we believe that a part of the reduction in extrapolated saturation moment is attributable to an antiferromagnetic component. It therefore appears that these compounds may be described as ferrimagnets. Careful neutron difftaction studies (which is possible for Tb compound) is urgently warranted to understand the complex magnetic structure.

The most fascinating finding which we are stressing in this article is that there are multiple steps in the M(H) curve at 1.8 K for Tb sample: If the sample is cooled very fast (within a few minutes) from 300 K to 1.8 K, there are three steps (near 6.2, 17.8, and 52 kOe in the specimen employed here) in the virgin curve in the data collected with VSM with a field-sweep rate of 4 kOe/min (figure 6). These steps do not appear when the field is reversed to zero, which we attribute to a 'supercooling' effect resulting in a phase-coexistence phenomenon at zero field after this field-cycling, as discussed in detail, for instance, for $Gd_5Ge_4$ [Ref. 11] and $Nd_7Rh_3$ [Ref. 12]. Further experiments are required, however, whether such a supercooling arises from magnetostructural effects or it is purely magnetic in origin. If the current to the magnet is reversed, two steps appear at -14.2 and -42.5 kOe. These steps vanish in the forward cycle when the field is reduced towards zero, but when the field enters positive quadrant, these reappear at 14.5 and 42.5 kOe. Interestingly, the transition field values are found to be marginally specimen-dependent, which we believe arises from partial occupancy of one of the Co sites, thereby resulting local environmental effects. We have also noted that, if the cooling rate is slow, these transitions in the virgin curve tend to broaden, similar to that seen for $Nd_7Rh_3$ [Ref. 12]. In addition, we have probed the behavior of these steps by measurements with the SQUID magnetometer and we find that the transitions occur at marginally higher fields (shifted by about 5 kOe) when compared to those obtained with VSM. The SQUID magnetometer measures magnetization after the stabilizing the field at a particular value (that is, the field-stable mode rather than field-sweep mode). Thus, it appears that the experimental conditions have a profound influence on these transitions in M(H). Thus, this compound exhibits interesting multiple metamagnetic transitions at 1.8 K. The hysteresis behavior at 1.8 K is distinctly different from that at a marginally higher temperature, say 5 K, and the irreversibility is observed even at fields as high as 80 kOe. There is a knee in the virgin curve in the M(H) curve, say, below 5 K, at low fields, which we attribute to the gradual dominance of an antiferromagnetic component and associated field-induced spin re-orientation effects, with decreasing temperature.

We have derived information about the MCE behavior from the isothermal M data, employing its relationship with isothermal entropy change ($\Delta S$) through Maxwell's equation [13]. The results of $\Delta S$ thus obtained are shown in figures 7 and 8 for a variation of the field from zero to a desired value. The values are maximum at the respective magnetic ordering temperatures. Since the magnetic ordering sets in near 300 K for Gd compound, it is quite tempting to compare the value with that of Gd for potential applications, particularly noting that the isothermal magnetization curves are not hysteretic for the Gd compound – a strongly desired factor for applications [14]. The value, say for H= 0 to 50 kOe, at the peak is close to half (about -6.5 J/mol-K or -42 mJ/cc-K) of that for Gd metal (-77 mJ/ccK, Ref. 13). For the Tb sample, the value at the peak is about -5.4 J/mol-K (in other words, -37 mJ/ccK), which is much lower compared to that (-210 mJ/ccK) of $LaFe_{11.7}Si_{1.3}$ ordering magnetically nearly at the same temperature. The refrigeration capacity (RC), defined as in Ref. 13, is found to be about 62 and 70 mJ/cc-kOe for the Gd and Tb compounds respectively, and these values are not negligible, considering that the corresponding values [13] for Gd metal and $LaFe_{11.7}Si_{1.3}$ are 110 and 93 mJ/cc-kOe. Our results on the Gd compound agree quite well with Ref. 10, which appeared in print after this article was communicated for publication. The negative sign of ΔS implies [13] that there is a dominating ferromagnetic component at high fields, at least above 20 K. However, there is a sign reversal in the ΔS data for Tb sample below 20 K with a significant magnitude at lower temperatures, the origin of which is not clear; possibly, this arises from a dominating antiferromagnetic component (arising from ferrimagnetism) persisting even at high fields and/or a small thermal/field cycling effects below this temperature can possibly have a profound effect on the M data at low temperatures. It is also known that ΔS derived from magnetization in a non-equilibrium state [15] may not be reliable and it is possible that this temperature marks the onset of such a non-equilibrium state.

It is now recognized theoretically [16, 17] that there is a relationship between $\Delta S$ and $H$. For instance, for magnetic materials with second order phase transition [17], $\Delta S = - kM_s(0)h^{2/3} - S(0,0)$, where $h$ is the reduced field (given by $\mu_0\mu_BH/k_BT_C$). $k$ is a constant and $M_s(0)$ is the saturation magnetization at low temperatures. The $\Delta S$ data of many materials were fitted to this equation successfully, but the value of $S(0,0)$ ranges from -0.2 to -1.06/kgK, the physical significance of which is not yet clear. We have therefore analysed the peak value of $\Delta S$ as a function of $h^{2/3}$ and plotted in the insets of figure 7 and 8. To enable a straightforward comparison with Ref.17, in the insets, we present $\Delta S$ in the units of J/kgK. It is found that there is a small deviation at low $h$ values for the Tb case as in Ref.17, but the plot is exceptionally linear for the Gd compound. The values of $S(0,0)$ and the coefficient of $h^{2/3}$ term for the Gd (Tb) compound are found to be -0.3 (-1.25 ) and -120 (-90) J/kgK very close to the values reported in table 1 of Ref. 13. The corresponding values of $k$ are about 0.6 and 0.46 respectively.

4. Summary

The magnetization behavior of the compounds, $Gd_6Co_{1.67}Si_3$ and $Tb_6Co_{1.67}Si_3$, ordering magnetically below about 300 K and 190 K has been systematically studied. The magnetic behaviors of these two compounds are found to be qualitatively different. The most notable of all the findings reported here is that, while increasing the magnetic field, we observe multiple steps in the isothermal magnetization for the Tb sample, that too at 1.8 K only; these steps are not observed when the field is reversed to zero and therefore this finding may have some relevance to the concept of 'super-cooling' and 'phase co-existence' following first order transitions [see, for instance, Refs. 11 and 12]. Such features are absent for the Gd compound. There is no hysteresis in the isothermal magnetization of the Gd compound, whereas, in the Tb compound, hysteresis loops are found at all temperatures, though these loops become smaller with increasing temperature. These compounds exhibit modest magnetocaloric effect at respective magnetic ordering temperatures and it is found that MCE varies essentially as $H^{2/3}$, rendering a support to the theory of Ref. 17.

References:
*Corresponding author: sampath@mailhost.tifr.res.in
1. B. Chevalier, E. Gaudin, and F. Weill, J. Alloys and Compounds **442,** 149 (2007).
2. E. Gaudin, S. Rence, and B. Chevalier, Solid State Sciences, **10,** 481 (2008).
3. E. Gaudin and B. Chevalier, J. Solid State Chem. **180**, 1397 (2007).
4. E. Gaudin, F. Weill, and B. Chevalier, Z. Naturforsch. **61b,** 825 (2006).
5. E. Gaudin, S. Tence, F. Weill, J. R. Fernandez, and B. Chevalier, Chem. Mater. **20,** 2972 (2008).
6. Niharika Mohapatra, Kartik K Iyer, and E.V. Sampathkumaran, Cond-mat arXiv:0712.1999.
7. S.B. Roy, A.K. Pradhan, P. Chaddah, and E.V. Sampathkumaran, J. Phys.: Condens. Matter**. 9**, 2465 (1997).
8. R. Welter, G. Venturini, E. Ressouche, and B. Malaman, J. Alloys and Comp. **210**, 279 (1994).
9. V.O. Garlea, J.L. Zarestky, C.Y. Jones, L.-L. Lin, D.L. Schlagel, T.A. Lograsso, A.O. Tsokol, V.K. Pecharsky, K.A. Gschneidner, Jr., and C. Stassis, Phys. Rev. B **72**, 104431 (2005).
10. Shen Jun, Li Yang-Xian, Dong Qiao-Yan, Wang Fang, and Sun Ji-Rong, Chin. Phys. Soc. **17**, 2268 (2008).
11. See, for instance, E.M. Levin, K.A. Gschneidner. Jr., and V.K. Pecharsky, Phys. Rev. B **65,** 214427 (2002). M.K. Chattopadhyay, M.A. Manekar, A.O. Pecharsky, V.K. Pecharsky, K.A. Gschneidner. Jr, G.K. Perkins, Y.V. Bugoslavsky, S.B. Roy, P. Chaddah, and L.F. Cohen, Phys. Rev. B **70,** 214421 (2004) and references therein.
12. Kausik Sengupta and E.V. Sampathkumaran, Phys. Rev. B 73, 020406R (2006).
13. See, for a review, A.M. Tishin, J. Magn. Magn. Mater. **316,** 351 (2007).
14. V. Provenzano, A.J. Shapiro, and R.D.Shull, Nature **429**, 853 (2004).
15. F. Casanova, A. Labarta, and X. Batlle, Phys. Rev. B **72,** 172402 (2005).
16. V. Franco, A. Conde, V.K. Pecharsky, and K.A. Gschneidner Jr. Europhys. Lett. **79**, 47009 (2007).
17. Qiao-yan. Dong, Hong-wei Zhang, Jue-lian Shen,Ji-rong Sun, and Bao-gen Shen, J. Magn. Magn. Mater. **319**, 56 (2007).

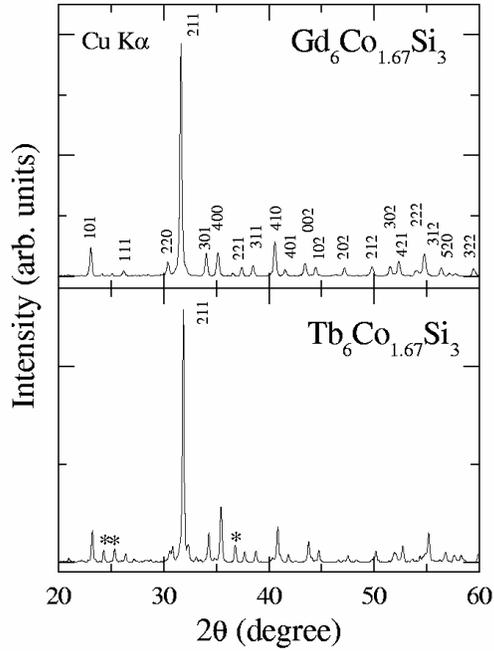

Figure1:
(color online) X-ray diffraction patterns for $Gd_6Co_{1.67}Si_3$ and $Tb_6Co_{1.67}Si_3$. Weak extra lines marked by asterisks are attributable to $Tb_5Si_3$.

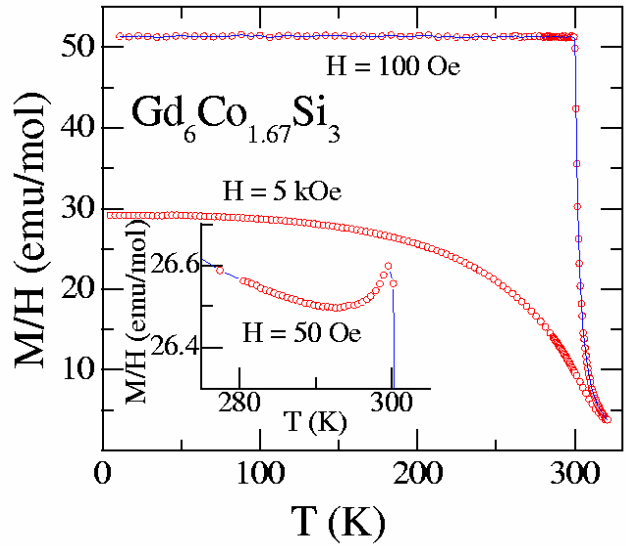

Figure 2:
(color online) Magnetization divided by magnetic field as a function of temperature, measured in the presence of various fields for $Gd_6Co_{1.67}Si_3$, for the zero-field-cooled (points) and field-cooled (lines) conditions of the specimen. In the inset, the data for H= 50 Oe is shown in an expanded form in the vicinity of the onset of magnetic transition for Gd sample.

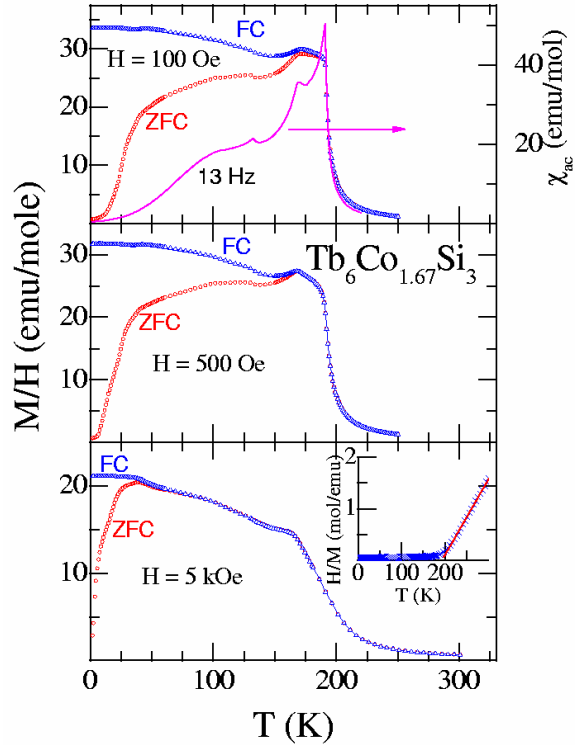

Figure 3:
(color online) Magnetization divided by magnetic field as a function of temperature, measured in the presence of various fields for $Tb_6Co_{1.67}Si_3$, for the zero-field-cooled and field-cooled conditions of the specimen. Ac susceptibility data measured with a frequency of 13 Hz is also plotted. In the inset, inverse susceptibility is plotted and a line through the Curie-Weiss region is drawn.

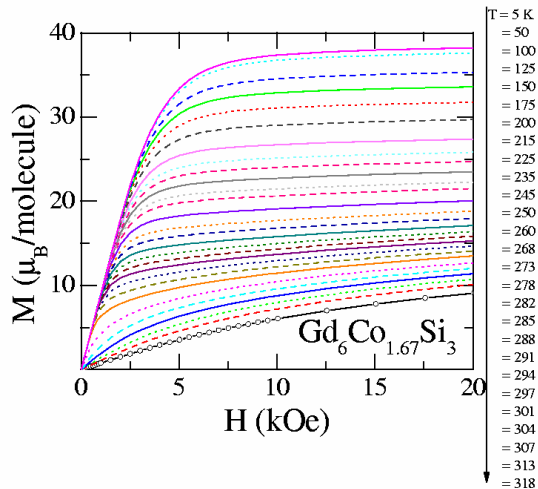

Figure 4:
(color online) Isothermal magnetization behavior at several temperatures for the zero-field-cooled condition of the specimen, $Gd_6Co_{1.67}Si_3$. The curves are non-hysteretic. While for 318 K, the data points are shown, for others, the lines only are shown for the sake of clarity.

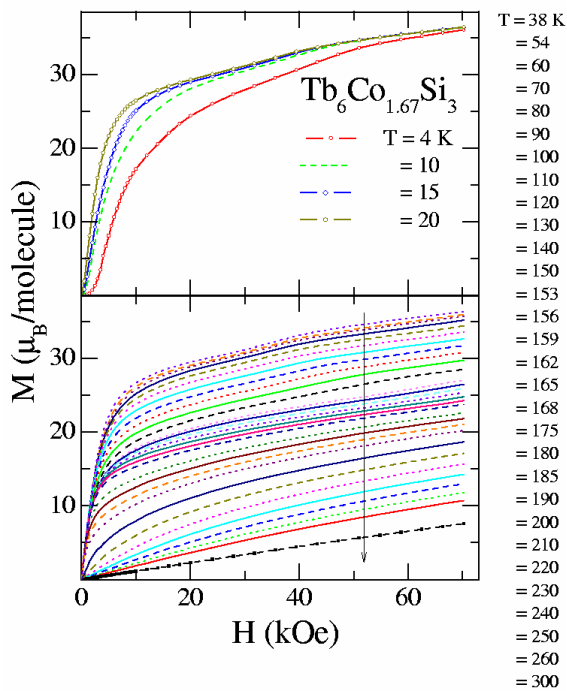

Figure 5:
(color online) Isothermal magnetization behavior (virgin curves) at several temperatures for the zero-field-cooled condition of the specimens, $Tb_6Co_{1.67}Si_3$, while increasing the field. The temperatures mentioned outside the figure are for the bottom portion of the figure.

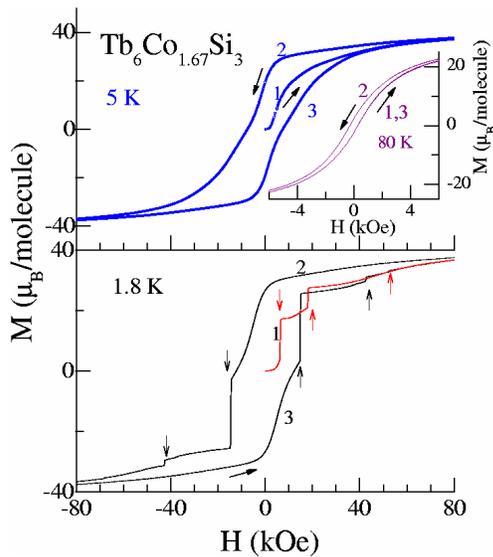

Figure 6:

(color online) Magnetic hysteresis loops for $Tb_6Co_{1.67}Si_3$ at 1.8 K, 5 K, and 80 K for the zero-field-cooled condition of the specimen measured with vibrating sample magnetometer with field-sweep mode. Vertical arrows for T= 1.8 K mark the magnetic fields at which discontinuous changes in magnetization are observed. Other arrows and numericals.(1, 2, 3) placed near the curves show the direction in which the field is changed.

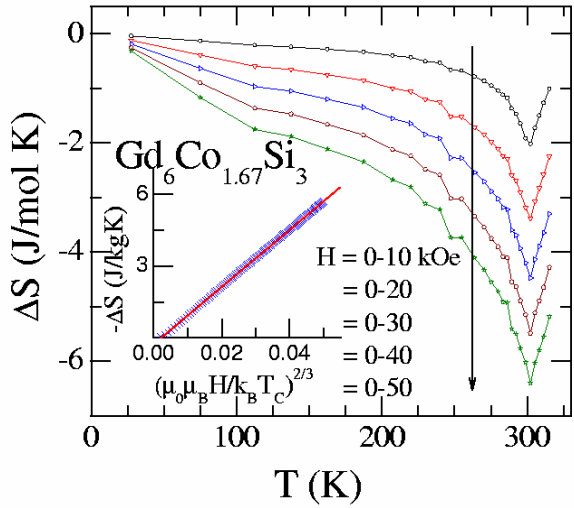

Figure 7:

(color online) Isothermal entropy change (a measure of magnetocaloric effect) for selected final magnetic fields (from the initial field of 0 Oe) as a function of temperature for $Gd_6Co_{1.67}Si_3$. A line is drawn through the data points and a vertical line is drawn to show the curves corresponding to final fields. In the inset, the same data (in different units) is plotted to show a functional dependence on magnetic field as described in the text and the line shows the linear region.

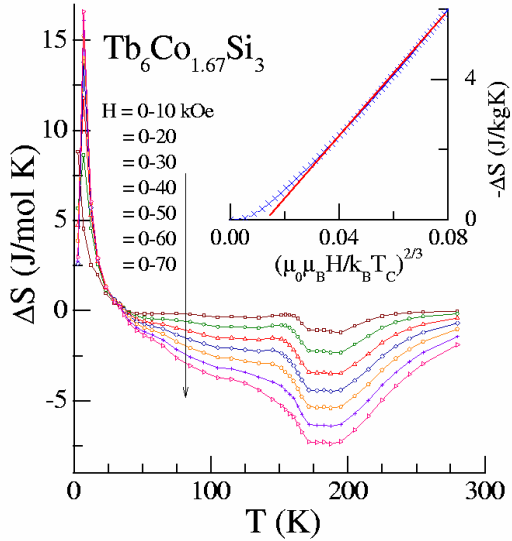

Figure 8:

(color online) Magnetocaloric data for $Tb_6Co_{1.67}Si_3$ as in figure caption 7.